\documentclass[%
 reprint,
superscriptaddress,
 amsmath,amssymb,
 aps,
 prl,
]{revtex4-1}

\usepackage{graphicx}
\usepackage{dcolumn}
\usepackage{bm}
\usepackage{hyperref}
\usepackage{color}

\begin{document}


\title{Angular momentum conservation in counter-propagating vectorially structured light}

\author{Hang Li}
\affiliation{
	School of Optical and Electronic Information, Huazhong University of Science and Technology, Wuhan, China} 
\author{Valeria Rodriguez-Fajardo}%

\affiliation{School of Physics, University of the Witwatersrand, Johannesburg, South Africa}
\author{Peifeng Chen}
\affiliation{
	School of Optical and Electronic Information, Huazhong University of Science and Technology, Wuhan, China} 
\author{Andrew Forbes}
\affiliation{School of Physics, University of the Witwatersrand, Johannesburg, South Africa}


\date{\today}

\begin{abstract}
\noindent It is well-known that electric spin angular momentum and  electric orbital angular momentum are conserved under paraxial propagation of travelling waves in free-space.  Here we study the electric and magnetic angular momentum in counter-propagating waves and show both theoretically and experimentally that neither component alone is conserved except in special cases.  We attribute this non-conservation to spin-spin and orbit-orbit coupling between the electric and magnetic fields. This work generalises previous findings based on travelling waves, explains the apparent spin-orbit coupling in counter-propagating paraxial light, and broadens our understanding of angular momentum conservation in arbitrary structured light waves.
\end{abstract}

\maketitle

\noindent The angular momentum (AM) of light has contributions from both the spin angular momentum (SAM) and the orbital angular momentum (OAM) components.  The former is dependent on the polarization of the light, with a purely left- and right-handed polarized beam carrying SAM of $\hbar$ and $-\hbar$ per photon, respectively.  The OAM is related to the spatial structure of the beam, specifically, a vortex phase of topological charge $\ell$ results in OAM of $\ell \hbar$ per photon \cite{allen1992orbital,PhysRevLett.88.053601,Yao:11}. Over the past few years, there has been enormous interest in the interactions between SAM and OAM due to its fundamental importance \cite{marrucci2011spin,Jisha:17,Zhu:14} and emerging applications in nanophotonics  \cite{PhysRevA.82.063825,cardano2015spin,padgett2011tweezers}. Spin-orbit (SO) interaction often occurs in non-paraxial beams including evanescent waves \cite{petersen2014chiral,o2014spin}, scattering \cite{PhysRevLett.102.123903,PhysRevLett.103.103903,PhysRevB.99.075155}, and tight focusing processes \cite{PhysRevA.97.053802,PhysRevA.97.053842}. Recently, SO interaction was reported in counter-propagating paraxial beams in free-space \cite{otte2018entanglement}, but could not explain why the total electric AM was not conserved.  In this paper, we propose that the non-conservation of electric angular momentum, including SAM, OAM and total AM, is a general phenomenon in counter-propagating beams. We attribute this phenomenon to spin-spin and orbit-orbit interaction between the electric and magnetic fields.  Previously it has been shown that the electric and magnetic fields contribute equally in travelling paraxial beams \cite{aiello2015note}, but now we make clear that while this is true, it is only so for travelling waves; counter-propagating beams do not obey this rule. We support our hypothesis by a theoretical framework that is validated by experiment, and offer a new perspective on spin-spin and orbit-orbit coupling in propagating light. 

\vspace{0.2cm}
\noindent \textbf{Theory.}  To begin, we consider a monochromatic electromagnetic field in free-space; by adopting the so-called
electric-magnetic democracy formalism \cite{berry2009optical}, the SAM density $\mathbf{s}$ and OAM density $\mathbf{l}$  can be written as \cite{aiello2015transverse} 
\begin{equation}\label{Eq:SAM &OAM}
\begin{aligned}
\mathbf{s}&=\frac{1}{4\omega} \mathrm{Im}[\epsilon_0 \mathbf{E}^*\times \mathbf{E}+\mu_0 \mathbf{H}^*\times \mathbf{H}],\\
\mathbf{l}&=\mathbf{r}\times\mathbf{p_o},
\end{aligned}
\end{equation}
\noindent for a radius vector of $\mathbf{r}$, with the orbital linear momentum density $\mathbf{p_o}$ defined as
\begin{equation}\label{Eq:OFD}
\mathbf{p_o}=\frac{1}{4\omega} \mathrm{Im}[\epsilon_0 \mathbf{E}^*\cdot(\nabla) \mathbf{E}+\mu_0 \mathbf{H}^*\cdot(\nabla) \mathbf{H}],
\end{equation}
\noindent and the spin linear momentum density as
\begin{equation}\label{SFD}
\mathbf{p_s}=\frac{1}{2}\nabla \times\mathbf{s},
\end{equation} 
\noindent for an angular frequency of $\omega$.  For compactness we have used the notation $\mathbf{A}\cdot (\nabla)\mathbf{B}=A_x\nabla B_x+A_y\nabla B_y+A_z\nabla B_z$, and all other terms have their usual meaning.  The total linear momentum is found from the sum of the orbital and spin contributions, $\mathbf{p}=\mathbf{p_o}+\mathbf{p_s}$.  Since the total linear momentum is proportional to the time-averaged Poynting vector (energy flow density), $\mathbf{p_o}$ and $\mathbf{p_s}$ can be regarded as the orbital and spin parts of this energy flow density: orbital flow density (OFD) and spin flow density (SFD), respectively  \cite{bekshaev2011internal,bekshaev2007transverse}.  We see from Eqs.~(\ref{Eq:SAM &OAM})-(\ref{SFD}) that the SAM, OAM, OFD and SFD consist of two separate terms: the first is only dependent on the electric field while the second term is only dependent on the magnetic field. Therefore, in the following, we will use superscripts $e$ and $m$ to distinguish between these electric and magnetic contributions.  Now we ask if there is coupling between these terms in counter-propagating light.  

In a travelling wave, the magnitudes of the electric and magnetic fields are proportional, so the spin and orbital angular momentum contributions by the electric and magnetic fields are equal  \cite{berry2009optical}.  However, the situation changes in counter-propagating beams. To illustrate this, we consider a general counter-propagating beam in which the slow dependence of the electromagnetic field on $z$ is neglected (we assume without loss of generality that the beams are collimated). In cylindrical coordinates this can be expressed in a general form as
\begin{equation}\label{Eq:ele}
\begin{aligned}
\mathbf{E}_\perp &=\mathbf{u}^+(r,\phi)\exp(ikz)+\mathbf{u}^-(r,\phi)\exp(-ikz),\\
E_z&=\frac{i}{k}\nabla_{\perp}\cdot[\mathbf{u}^+\exp(ikz)-\mathbf{u}^-\exp(-ikz)],
\end{aligned}
\end{equation}
\noindent where $\nabla_\perp=\mathbf{e}_x\partial x+\mathbf{e}_y\partial y$, $k=\omega/c$ is the wavenumber, and $c$ is the speed of light.  The total electric field is $\mathbf{E}=\mathbf{E}_\perp+E_z\mathbf{e}_z$, in which $E_z$ is obtained by $\nabla \cdot \mathbf{E}=0$.  The first term with $\exp(ikz)$ is the part whose direction of propagation is parallel to the unit vector $\mathbf{e}_z$, while the second term with $\exp(-ikz)$ is antiparallel to $\mathbf{e}_z$.  In this notation, $\mathbf{u}^+$ and $\mathbf{u}^-$ are positive- and negative-propagating modes, respectively, and can be any structured light field:   Laguerre-Gaussian, Hermite-Gaussian, cylindrical vector vortex beams, and so on.  Since the wavenumber, $k$, is very large, the longitudinal component of the electric field is much smaller than transverse component.  Small but not negligible, as we will show that the longitudinal electromagnetic field plays a significant role in transverse SFD in counter-propagating beams.  Using the first-order approximation \cite{bekshaev2011internal}, the magnetic field $\mathbf{H}=\mathbf{H}_\perp+H_z\mathbf{e}_z$ is 
\begin{equation}\label{Eq:mag}
\begin{aligned}
\mathbf{H}_\perp&=\frac{1}{c\mu_0}\mathbf{e}_z\times[\mathbf{u}^+\exp(ikz)-\mathbf{u}^-\exp(-ikz)],\\
H_z&=\frac{i}{\omega\mu_0}\nabla_\perp\cdot(\mathbf{e}_z\times\mathbf{E}_\perp).
\end{aligned}
\end{equation}
It should be noted that there is an extra minus sign in the second term of the transverse magnetic field. If the two parts of transverse electric field are in phase, the two parts of transverse magnetic field will be exactly out of phase, which means  they  contribute unequally to the momentum.

In paraxial beams, only the $z$-component of the SAM and OAM is of primary relevance: $S_z$ and $L_z$, respectively.  They are the sum of the electric and magnetic contributions, $S_z=S_z^e+S_z^m$ and $L_z=L_z^e+L_z^m$,  and found from the SAM and OAM densities 
\begin{equation}\label{Eq:global_SAM&OAM}
\begin{aligned}
S_z^e&=\frac{\int\int  s_z^e r\mathrm{d}r\mathrm{d}\phi}{\tfrac{1}{\hbar\omega}\int\int w r\mathrm{d}r\mathrm{d}\phi },
S_z^m&=\frac{\int\int s_z^m r\mathrm{d}r\mathrm{d}\phi}{\tfrac{1}{\hbar\omega}\int\int wr\mathrm{d}r\mathrm{d}\phi }, \\
L_z^e&= \frac{\int\int l_z^e  r\mathrm{d}r\mathrm{d}\phi}{\tfrac{1}{\hbar\omega}\int\int wr\mathrm{d}r\mathrm{d}\phi },
L_z^m&=\frac{\int\int l_z^m   r\mathrm{d}r\mathrm{d}\phi}{\tfrac{1}{\hbar\omega}\int\int wr\mathrm{d}r\mathrm{d}\phi },
\end{aligned}
\end{equation}
with the energy density $w$ defined as
\begin{equation}\label{Eq:enrgy}
\begin{aligned}
w&={\textstyle\frac{1}{4}}\left(\epsilon_0 \mathbf{E}^*\cdot \mathbf{E}+\mu_0 \mathbf{H}^*\cdot \mathbf{H}\right) \\&= {\textstyle\frac{1}{2}}\epsilon_0\left(|\mathbf{u}^+|^2+|\mathbf{u}^-|^2\right).
\end{aligned}
\end{equation}
In Eq.~(\ref{Eq:global_SAM&OAM}), $s_z^e$, $s_z^m$, $l_z^e$ and $l_z^m$ are the longitudinal electric SAM density, magnetic SAM density, electric OAM density and magnetic OAM density, respectively. The numerators in Eq.~(\ref{Eq:global_SAM&OAM}) are the global SAM/OAM while the denominators represent the number of photons.  Equation (\ref{Eq:enrgy}) shows that the energy density $w$ is independent of $z$, which means the number of photons (or the energy) is conserved. Therefore, whether the angular momenta per photon ($S_z^e$, $S_z^m$, $L_z^e$ and $L_z^m$) are conserved is equivalent to whether the global angular momenta are conserved.

After a little algebra (see Supplementary Information), one finds
\begin{small}
	\begin{equation}
	\begin{aligned}
	s_z^e \mathbf{e}_z &\propto \mathbf{E}_\perp^*\times\mathbf{E}_\perp= \mathbf{u}^{+*}\times \mathbf{u}^++\mathbf{u}^{-*}\times \mathbf{u}^-\\
	&\quad  +\mathbf{u}^{+*}\times\mathbf{u}^-\exp(-i2kz)+\mathbf{u}^{-*}\times\mathbf{u}^+\exp(i2kz)  \\
	S_z&=\frac{\hbar\mathrm{Im}\left\{\int\int \mathbf{e}_z\cdot r(\mathbf{u}^{+*}\times \mathbf{u}^++\mathbf{u}^{-*}\times \mathbf{u}^-)\mathrm{d}r\mathrm{d}\phi\right\}}{\int\int (|\mathbf{u}^+|^2+|\mathbf{u}^-|^2)r\mathrm{d}r\mathrm{d}\phi } \\
	L_z&=\frac{\hbar\mathrm{Im}\left\{\int\int (\mathbf{u}^{+*}\cdot\partial_\phi \mathbf{u}^++\mathbf{u}^{-*}\cdot\partial_\phi \mathbf{u}^-)r\mathrm{d}r\mathrm{d}\phi\right\}}{\int\int (|\mathbf{u}^+|^2+|\mathbf{u}^-|^2)r\mathrm{d}r\mathrm{d}\phi }.
	\label{Eq:global_ AM}
	\end{aligned}
	\end{equation}
\end{small}
\noindent The first two terms of $\mathbf{E}_\perp^*\times\mathbf{E}_\perp$ give the contribution of the positive- and negative-propagating parts of the electric SAM, respectively. The last two terms are cross contributions, which are not zero except in some special cases.  We conclude that the electric SAM is not conserved, which seems contrary to intuition. The total OAM and SAM are both independent of $z$, so they are conserved.   In the special case that $\mathbf{u}^-=0$, the field degenerates into a one-way travelling beam; the cross contribution vanishes and the electric SAM is again conserved, consistent with intuition and prior reports \cite{aiello2015note}. Similar conclusions can be also reached on the magnetic SAM, electric OAM and magnetic OAM (see Supplementary Information).

\vspace{0.2cm}
\noindent \textbf{Example.} To make this practical for testing, consider a vector beam that is composed of a superposition of two orthogonally polarized Laguerre-Gaussian modes, $\text{LG}_{p,\ell} (r,\phi)$ (carrying OAM of $\ell \hbar$ per photon) given by
\begin{equation}
\begin{aligned}
\mathbf{u}^+&={\textstyle\frac{1}{2}}[\exp(i\alpha)\text{LG}_{p_1,\ell_1}\mathbf{e_R}+\exp(-i\alpha)\text{LG}_{p_2,\ell_2}\mathbf{e_L}], \\
\mathbf{u}^-&={\textstyle\frac{1}{2}}[\exp(i\beta)\text{LG}_{p_1,\ell_1}\mathbf{e_R}+\exp(-i\beta)\text{LG}_{p_2,\ell_2}\mathbf{e_L}],
\label{Eq:3u}
\end{aligned}
\end{equation}

\noindent where $\mathbf{e_R}=(\mathbf{e_x}-i\mathbf{e_y})/\sqrt{2}$ and $\mathbf{e_L}=(\mathbf{e_x}+i\mathbf{e_y})/\sqrt{2}$ represent the unit vector of right- and left-circularly polarized states, respectively, and $\alpha$ and $\beta$ define the phase relation between the two states. When $\alpha=\beta$, the polarization states of the positive-propagating and negative-propagating components are the same. When $|\alpha-\beta|=\pi/2$, their states are orthogonal. 
In Eq.~(\ref{Eq:3u}), the weight factors of the right- and left-circularly polarized components are equal (for simplicity, but can easily be generalized). The SAM and OAM per photon carried by the right-circularly polarized component is $-\hbar$ and $\ell_1 \hbar$, respectively, and $\hbar$ and $\ell_2 \hbar$ for the left-circularly polarized component. Hence, the total SAM and OAM per photon are 0 and $(\ell_1+\ell_2)\hbar/2$, respectively.
\begin{figure}
	\centering
	\includegraphics[width=8.6cm]{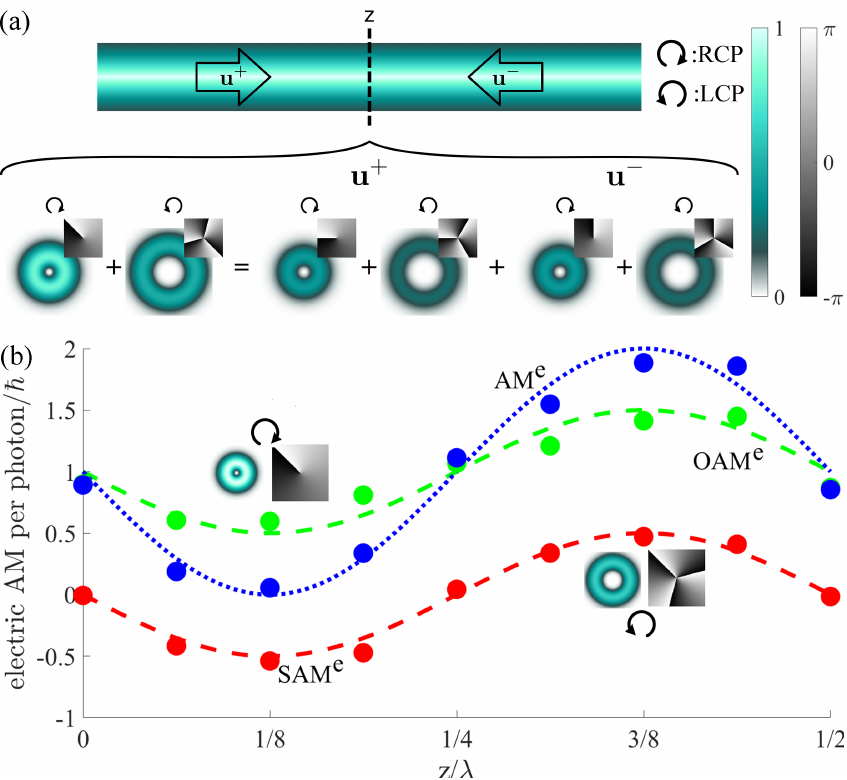}
	\caption{ (a) Conceptual sketch of the counter-propagating beam, with intensity and phase distributions for $p_1=p_2=0$, $\alpha=0,\beta=\pi/2$ and $\ell_1=1$ and $\ell_2=3$ at $z=0$ shown.  (b) Simulated (lines) and experimental (points) results of global electric SAM, OAM and total AM per photon. The insets show the intensity and phase distributions of the field at $z=\lambda/8$ (trough) and $z=3\lambda/8$ (peak). Error bars are too small to be seen.}\label{Fig1}
\end{figure}

To test the theory, we engineer our modes with no radial content ($p_1=p_2=0$) since global SAM and OAM are independent of radial modes, a Rayleigh length of $\approx 6$ m so that the fields can be regarded as collimated, and set $\alpha=0,\beta=\pi/2$ and $\ell_1=1$ and $\ell_2=3$, shown schematically in Fig.~\ref{Fig1}(a), with full details of the experiment given in the Supplementary Information.  The electric SAM, OAM and AM per photon were measured and compared to theory, shown in Fig. \ref{Fig1}(b).   As predicted, they are all non-conserved qualities which increase and decrease in phase resulting in no global spin-orbit coupling (in this case).  The right- and left-circularly polarized components are both standing waves with nodes at $z= \lambda/8$ and $z=3\lambda /8$, respectively, where the superposition is a scalar mode (see Supplementary Information).  Moving along the axis from $z=\lambda/8$ to $z=3\lambda/8$, a purely $\text{LG}_{0,1}$ mode with right-circular polarization is gradually transformed into a left-circularly polarized $\text{LG}_{0,3}$ mode, causing both OAM and SAM to increase, with the process reversed from $z=3\lambda/8$ to $z=5\lambda/8$. The explanation for this non-conservative requires consideration of the magnetic field component. 
\begin{figure*}[htbp]
	\centering
	\includegraphics[width=17cm]{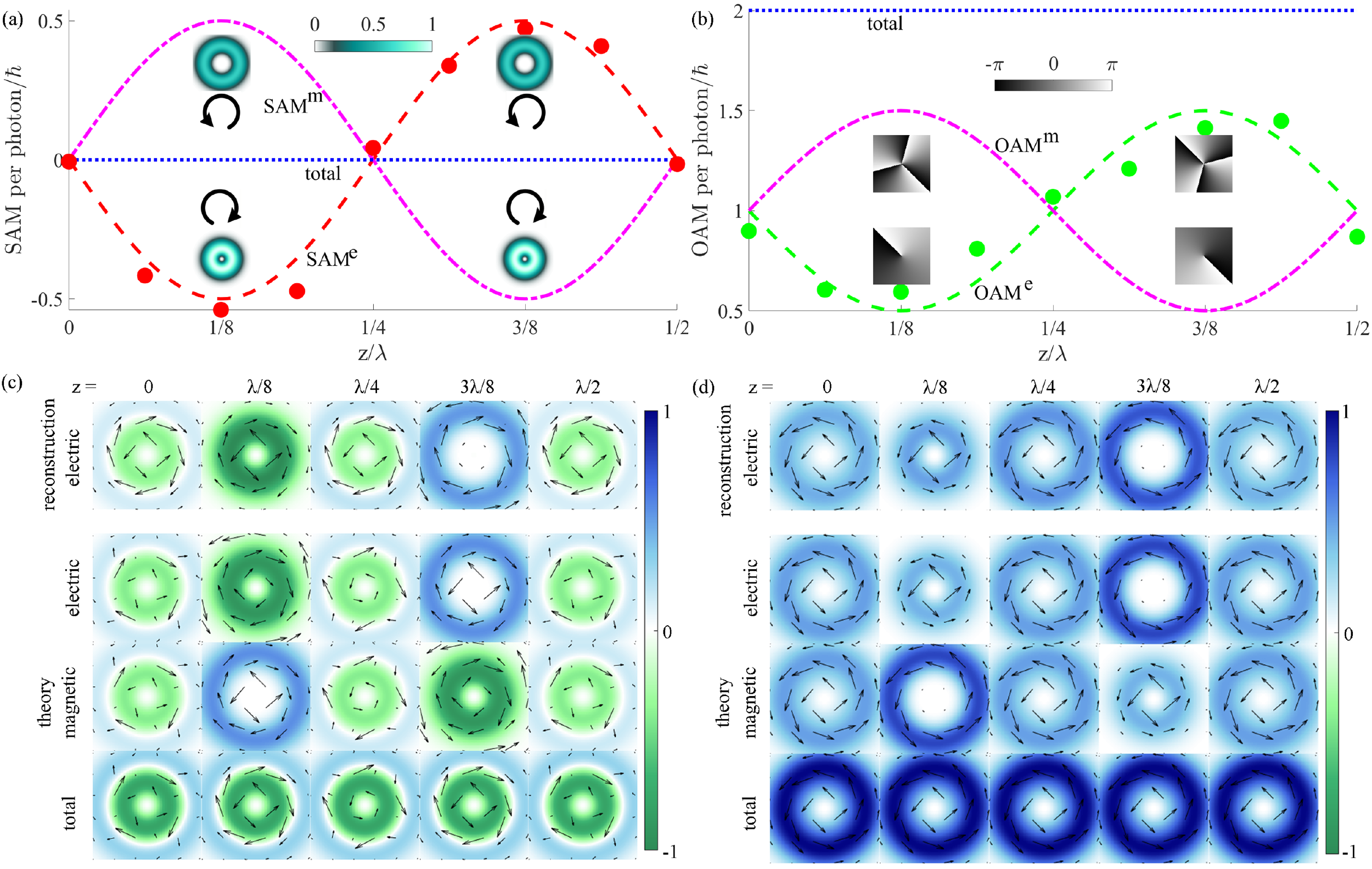}
	\caption{Spin-spin coupling and orbit-orbit coupling between the electric and magnetic fields. We show the electric, magnetic and total (a) SAM and (b) OAM per photon, theory (lines) and experimental electric data (points). In (c) and (d) we show the corresponding longitudinal SAM and OAM densities (false colour plots), respectively, overlaid with their transverse SFD and OFD (black arrows) in the XY-plane.  The first row of each shows the experimentally reconstructed densities and flows for the electric component, while the second, third and fourth show the simulations.}	\label{Fig2}
\end{figure*}

\vspace{0.2cm}
\noindent \textbf{Spin-spin coupling.}   Using the earlier analysis, the electric SAM and magnetic SAM per photon for our experimental example can be expressed as
\begin{equation}\label{Eq:3SAM}
\begin{aligned}
S_z^e&={\textstyle\frac{1}{2}}\hbar\sin(\alpha-\beta)\sin(2kz),\\
S_z^m&={\textstyle\frac{1}{2}}\hbar\sin(\beta-\alpha)\sin(2kz).
\end{aligned}
\end{equation}
Figure~\ref{Fig2} (a) shows the calculated electric, magnetic and total SAM per photon as a function of $z$, with experimental electric data in good agreement with theory.  We find that the electric and magnetic SAM vary while the total SAM is always zero, which we interpret as spin-spin coupling between the electric and magnetic fields.  This coupling results from the polarization variation in our vectorial beam: we have a standing wave in polarization structure rather than intensity.  According to Eq.~(\ref{Eq:3SAM}), the coupling will vanish when the polarization states of the positive-propagating part and the negative-propagating part are the same, that is, when $|\alpha-\beta|=0$ or $\pi$, because the polarization structure is then invariant along the $z$-axis. When their polarized states are orthogonal, that is, $|\alpha-\beta|=\pi/2$, the coupling will be the strongest. In other cases, the coupling still exists but varies in strength.

The longitudinal SAM density and transverse SFD (superimposed arrows) are shown in Fig.~\ref{Fig2} (c), with the experimentally reconstructed electrical component in the first row and the simulation in the second, third and fourth rows. We see that the electric longitudinal SAM density and transverse SFD from $z=0$ to $z=\lambda/2$ are the same as the magnetic part from $z=\lambda/2$ to $z=0$. The opposite variability of electric and magnetic SAM density causes invariant total SAM density, as shown in the last row of Fig.~\ref{Fig2} (c). The transverse electric SFD was calculated from 
\begin{equation}\label{transSFD}
\mathbf{p}_{s\perp}^e={\textstyle\frac{1}{2}}\left[\nabla_\perp s_z^e\times\mathbf{e}_z+\mathbf{e}_z\times\partial_z\mathbf{s}_\perp^e\right].
\end{equation}

In a one-way travelling collimated beam, the second term is always zero, but not so in our theoretical example. In our experimental measurement, we used a constructed counter-propagating beam, resulting in the disparity in the theoretical and experimental SFDs. The second term in Eq. (\ref{transSFD}) is dependent on the longitudinal electric field while the longitudinal electric SAM density is circularly symmetrical (because of the circular symmetry of LG modes), so the first term only has an azimuthal component.   Thus, the reconstructed experimental electric SFD, first row of Fig. \ref{Fig2} (c), is either clockwise or anticlockwise.  In counter-propagating beams, the second term is no longer zero because of the cross contribution from the positive-propagating and the negative-propagating beams.  This term makes the modes of counter-propagating beams vary between left-circularly polarized ($\text{LG}_{0,3}$) and right-circularly polarized ($\text{LG}_{0,1}$) modes. For example, from $z=\lambda/8$ to $z=3\lambda/8$, the electric mode is transformed from an $\text{LG}_{0,1}$ mode into a $\text{LG}_{0,3}$ mode, and vice versa for the magnetic mode.   

\vspace{0.2cm}
\noindent \textbf{Orbit-orbit coupling.} For our test example, the electric and magnetic OAM can be expressed as
\begin{small}
	\begin{equation}\label{Eq:3OAM}
	\begin{aligned}
	L_z^e&={\textstyle\frac{1}{2}}\hbar\left[l_1\cos^2(kz+\tfrac{\alpha-\beta}{2})+l_2\cos^2(kz+\tfrac{\beta-\alpha}{2})\right],\\
	L_z^m&={\textstyle\frac{1}{2}}\hbar\left[l_1\sin^2(kz+\tfrac{\alpha-\beta}{2})+l_2\sin^2(kz+\tfrac{\beta-\alpha}{2})\right].
	\end{aligned}
	\end{equation}
\end{small}
In Fig.~\ref{Fig2} (b) we show simulations for the electric, magnetic and total OAM per photon as function of $z$, with the measured electric OAM shown as data points, in excellent agreement with theory.  This confirms orbit-orbit coupling between the electric and magnetic fields. This coupling is not guaranteed, and instead is dependent on both the spin states and the (OAM) topological charges, $\ell_1$ and $\ell_2$.  There are three cases in which the coupling will vanish: (1) when  $\ell_1=\ell_2=0$, neither the right- nor the left-circularly polarized components of the beam have a vortex phase, leading to zero electric OAM and zero magnetic OAM, so no possibility of orbit-orbit coupling; (2) when $\alpha=\beta$ and $\ell_1= - \ell_2$, the polarization is structurally invariant along the $z$-axis and the OAM contributed by the two component cancel: both the electric and magnetic OAM will be zero; (3) when $|\alpha-\beta|=\pi/2$ and $\ell_1=\ell_2$, the spin states of the counter-propagating components are orthogonal, and the electric energy per unit length is conserved. Thus $\ell_1=\ell_2$ results in a balance between the electric OAM contributed by the two components - the electric and magnetic OAM are both conserved independently and the coupling appears to vanish.

Finally, the longitudinal OAM density and transverse OFD (superimposed arrows) are shown in Fig.~\ref{Fig2} (d), with the experimentally reconstructed electrical components (first row) in excellent agreement with theory (second row).  We see that the electric longitudinal OAM density and transverse OFD from $z=0$ to $z=\lambda/2$ are the same as the magnetic part from $z=\lambda/2$ to $z=0$. The opposite trend of electric and magnetic OAM density results in an invariant total OAM density, as shown in the last row of Fig. \ref{Fig2} (d). The transverse electric OFD,  $\mathbf{p}_{o\perp}^e=l_z^e\mathbf{e}_\phi/r$, has no radial component and is independent of the longitudinal electric field. Further, since the topological charges $\ell_1$ and $\ell_2$ are both positive, the transverse OFD (magnetic or electric) is always anticlockwise, resulting in a longitudinal OAM density that is always positive. If one topological charge is positive and the other is negative, OFD reversal occurs.

\vspace{0.1cm}
\noindent \textbf{Conclusion.} We have demonstrated that generally held conservation rules appear to be violated in counter-propagating vectorially structured light beams, which we confirm by experiment.  We reconcile this by considering terms that are traditionally held to be zero or `the same', and interpret the results as spin-spin and orbit-orbit coupling.  With this we are able to explain some anomalies previously reported, while reducing to the well-known results for travelling waves under special cases.  We find that in general the total longitudinal SAM and OAM are both conserved, while neither the electric part nor magnetic part are conserved independently.  This work generalises previous findings based on travelling waves, explains the apparent spin-orbit coupling in counter-propagating paraxial light, and broadens our understanding of angular momentum conservation in arbitrary structured light waves.

\end{document}